\documentclass[preprint,authoryear,12pt]{elsarticle}
\usepackage[utf8]{inputenc}

\usepackage{fullpage}

\usepackage{graphicx}%
\usepackage{multirow}%
\usepackage{multicol}%
\usepackage{amsmath,amssymb,amsfonts}%
\usepackage{amsthm}%
\usepackage{mathrsfs}%
\usepackage[title]{appendix}%
\usepackage{xcolor}%
\usepackage{textcomp}%
\usepackage{manyfoot}%
\usepackage{booktabs}%
\usepackage{listings}%
\usepackage{url}

\usepackage{algorithm,algorithmicx}
\usepackage[noend]{algpseudocode}

\usepackage[normalem]{ulem}
\usepackage{tikz}

\newcommand{\mint}{\mbox{\rm min}_t}
\newcommand{\maxt}{\mbox{\rm max}_t}
\newcommand{\real}{\mathbb{R}}

\newcommand{\prob}{HRCP}

\newtheorem{proposition}{Proposition}
\newtheorem{definition}{Definition}
\newtheorem{remark}{Remark}
\newtheorem{lemma}{Lemma}

\newcommand{\X}[0]{\mathcal{X}}


\journal{EJOR}

\begin{document}

\begin{frontmatter}

	\title{Exact Hyper-Rectangular Clustering via Adaptive Subset Selection}

\author[essec]{Diego Delle Donne\corref{cor1}}
\affiliation[essec]{
    organization={ESSEC Business School},
    city={Cergy-Pontoise},
    country={France}}
\ead{delledonne@essec.edu}
\cortext[cor1]{Corresponding author}

\author[utdt]{Javier Marenco}
\affiliation[utdu]{
    organization={Business School, Universidad Torcuato Di Tella},
    city={Buenos Aires},
    country={Argentina}}
\ead{javier.marenco@utdt.edu}

\author[google]{Eduardo Moreno}
\affiliation[google]{
    organization={Google Research, Google},
    city={Paris},
    country={France}}
\ead{eduardo.moreno@gmail.com}

\begin{abstract}
    We study the hyper-rectangular clustering problem (HRCP), where the goal is to partition a set of points into a fixed number of axis-aligned clusters while minimizing the total span. Existing exact approaches, based on mathematical optimization formulations, are limited to relatively small instances due to their strong dependence on the number of data points.
    We propose an incremental exact algorithm that exploits a key structural property of the problem: optimal cluster boundaries are determined by a subset of points, while interior points do not affect the objective value. The algorithm iteratively solves HRCP on carefully selected subsets of points and expands them only when necessary. A simple optimality condition ensures that, once a solution covers the entire dataset, it is optimal for the original problem.
    We introduce sampling strategies designed to identify points likely to lie on cluster boundaries, guiding the incremental process toward informative subsets. 
    Computational experiments show that the proposed approach significantly improves scalability, solving instances with up to 10,000 points and substantially outperforming monolithic formulations.
\end{abstract}

	
\end{frontmatter}

\section{Introduction}
\label{sec.introduction}

Machine learning techniques are increasingly used to extract structure and
interpretable patterns from large-scale datasets. In this context, clustering
methods that provide not only a partition of the data but also an explicit and
interpretable description of the clusters are particularly appealing.
Hyper-rectangular clustering is a natural model in this direction, as each cluster
is defined by axis-aligned bounds on the features, yielding a representation that
is both interpretable and easy to communicate. This property makes hyper-rectangular
clusterings especially suitable for applications where transparency and
explainability are essential.

The hyper-rectangular clustering problem (HRCP) has been studied using a variety
of mathematical optimization techniques, including mixed-integer programming,
branch-and-cut, and branch-and-price approaches. These methods have proven
effective in solving instances to optimality and in providing strong modeling
frameworks for explainable clustering. However, a major limitation remains: their
scalability is severely constrained by the number of data points. As a consequence,
existing exact approaches are typically restricted to relatively small instances,
which limits their applicability in real-world settings.

In this work, we exploit a fundamental structural property of hyper-rectangular
clusterings: optimal clusters are determined by a small subset of points that define
the boundaries of the corresponding hyper-rectangles, while interior points do not
affect the objective value. This observation suggests that solving HRCP on the full
dataset may be unnecessary, as only a subset of points is required to define an optimal solution. In particular, it raises the question of whether it is
possible to identify a subset of “representative” points that suffices to recover an
optimal solution.

Building on this insight, we propose an incremental exact algorithm that solves
HRCP on carefully selected subsets of points. Starting from a small subset, the
algorithm iteratively expands it only when needed. A key feature of the approach is
a simple yet powerful optimality condition: as soon as the solution obtained on a
subset covers the entire dataset, it is guaranteed to be optimal for the original
problem. This allows the algorithm to certify optimality while solving significantly
smaller instances, effectively bypassing the main scalability bottleneck of existing
approaches.

The effectiveness of the incremental framework critically depends on selecting
informative points. To this end, we introduce sampling strategies designed to
identify points that are likely to lie on the boundaries of clusters. These
strategies exploit local geometric properties of the data and guide the incremental
process toward subsets that capture the essential structure of the optimal solution.

Our contributions can be summarized as follows:
\begin{itemize}
\item We propose a novel incremental exact framework for HRCP that reduces the
dependence on the number of data points by solving a sequence of smaller subproblems.
\item We establish a theoretical condition under which a solution obtained on a
subset is guaranteed to be optimal for the full dataset.
\item We introduce sampling strategies tailored to identify boundary-defining points,
which significantly improve the efficiency of the incremental procedure.
\item We provide an extensive computational study showing that the proposed approach
substantially outperforms monolithic formulations and scales to instances with up to
10,000 points.
\end{itemize}

Computational results demonstrate that the incremental approach dramatically
improves scalability with respect to the number of points, while preserving
solution quality. In contrast, classical monolithic formulations quickly become
intractable as the dataset grows. These findings highlight that, for HRCP, solving
the full problem does not require considering all data points, but rather focusing on a small subset of boundary-defining points.

The remainder of this work is organized as follows. Section~\ref{sec:problem} formally presents HRCP and provides a review of related literature. Section~\ref{sec:algorithm} presents the extended formulation this work is based upon, and details the proposed branch-and-price procedure for tackling this formulation. Section~\ref{sec:experimentation} reports our computational experience with an implementation of this procedure, whereas Section~\ref{sec:conclusions} closes the paper with concluding remarks and lines for future research.

\section{Formalization of the problem}\label{sec:problem}

Given a nonempty set ${\cal X} = \{ x^1, \dots, x^n \}$ of $n$ points in $\mathbb{R}^d$ and an integer $p\ge 1$, a \emph{$p$-clustering} of ${\cal X}$ is a collection of subsets $C_1,\dots,C_p \subseteq {\cal X}$, in such a way that ${\cal X} = C_1 \cup \dots \cup C_p$ and $C_i \cap C_j = \emptyset$ for $i,j\in\{1,\dots,p\}$, $i\ne j$. Each set from $\{C_i\}_{i=1}^p$ is called a \emph{cluster} in this context. The \emph{span} of a cluster $C\subseteq {\cal X}$ over the coordinate $t$ is $\mbox{span}_t(C) = \max\{x_t: x\in C\} - \min\{x_t: x\in C\}$ if $C \ne \emptyset$ and $\mbox{span}_t(C) = 0$ otherwise, and the \emph{total span} of $C$ is $\mbox{span}(C) = \sum_{t=1}^d \mbox{span}_t(C)$. Finally, the \emph{total span} of a clustering $\mathbb{C} = \{C_1,\dots,C_p\}$ is defined as the sum of the total spans of its constituent clusters, i.e., $\mbox{span}(\mathbb{C}) = \sum_{i=1}^p \mbox{span}(C_i)$.
% Given a set ${\cal X}$ of points, the integer $p$ specifying the number of clusters, and an integer $q\ge 0$, 
The \emph{hyper-rectangular clustering problem with axis-parallel clusters} (\prob) consists in determining a $p$-clustering of ${\cal X}$, minimizing the total span. 

As we may find in \cite{marenco23}, \prob{} can be easily formulated by means of mixed integer programming models. 
For $i\in [n] = \{ 1, \dots, n\}$ and $c\in [p] = \{ 1, \dots, p\}$, we consider the binary variable $z_{ic}$ representing whether $x^i$ is assigned to the cluster $c$ or not. Also, for $c\in [p]$ and $t\in [d] = \{ 1, \dots, d\}$, the real variables $l_{tc}, r_{tc} \in \mathbb{R}$ represent a lower and an upper bound, respectively, for the points in the cluster $c$ in the coordinate $t$. 
%For $t\in [d]$, define $X_t := \{ x_t: x\in {\cal X}\}$, $\mint := \min(X_t)$, and $\maxt := \max(X_t)$. In this setting, we can formulate the problem as follows.
For $t\in [d]$, we define $\mint := \min\{ x_t: x\in {\cal X}\}$ and $\maxt := \max\{ x_t: x\in {\cal X}\}$. In this setting, we can formulate the problem as follows.
\begin{align}
\min \sum_{c=1}^p \sum_{t=1}^d r_{tc} & - l_{tc} \label{cm.constr.0} \\
\mbox{s.t.} \ \sum_{c=1}^p z_{ic} \ & = \ 1 && \forall i\in [n], \label{cm.constr.1} \\
l_{tc} + (\maxt - x^i_t) z_{ic} \ & \le \ \maxt && \forall i\in [n], c\in [p], t\in [d], \label{cm.constr.2} \\
r_{tc} + (\mint - x^i_t) z_{ic} \ & \ge \ \mint && \forall i\in [n], c\in [p], t\in [d], \label{cm.constr.3} \\
l_{tc} \ & \le \ r_{tc} && \forall c\in [p], t\in [d], \label{cm.constr.4} \\
% \sum_{c=1}^p \sum_{i=1}^n z_{ic} & \ge & n-q, \label{cm.constr.5} \\
\mint \ \le \ l_{tc}, r_{tc} \ & \le \ \maxt && \forall c\in [p], t\in [d], \label{cm.constr.6} \\
z_{ic} \ & \in \ \{ 0, 1 \} && \forall i\in [n], c\in [p]. \label{cm.constr.7}
\end{align}  
The objective function asks to minimize the sum of the total cluster spans. Constraints (\ref{cm.constr.1}) ask every point to be assigned to one cluster. Constraints (\ref{cm.constr.2})-(\ref{cm.constr.3}) bind the variables, in such a way that $l_{tc} \le x^i_t \le r_{tc}$ if the point $i$ is assigned to the cluster $c$. 
Constraints (\ref{cm.constr.4}) avoid bound crossings in empty clusters, whereas constraints (\ref{cm.constr.6}) impose bounds for the $l$- and the $r$-variables. Finally, constraints (\ref{cm.constr.7}) specify that the $z$-variables are binary.

The formulation (\ref{cm.constr.0})-(\ref{cm.constr.7}) has obvious symmetry issues (i.e., every clustering admits more than one representation within the model, by renaming the cluster indices), which could be problematic when attempting the solution of this model with general integer programming solvers.
This can be tackled by adding symmetry-breaking constraints. Unfortunately, as reported in \cite{marenco23}, the addition of these does not appear to be effective in practice. \\

Alternatively, the same problem can be formulated as a constraint programming model on the same set of variables. In this setting, the following formulation models HRCP.
\begin{align}
    \min \sum_{c=1}^p \sum_{t=1}^d r_{tc} & - l_{tc} \label{eq:cp:begin} \\
	\mbox{s.t.} \ z_{ic} \ \Rightarrow \ l_{tc} \ & \leq \ x_t^i  && \forall i\in[n], c\in[p], t\in[d], \label{eq:cp:constr1} \\
	z_{ic} \ \Rightarrow \ r_{tc} \ & \geq \ x_t^i && \forall i\in[n], c\in[p], t\in[d], \label{eq:cp:constr2} \\
	\textsc{ExactlyOne}(z_{i1},& \ldots,z_{ip}) &&  \forall i\in[n], \label{eq:cp:exactlyone} \\
    \mint \ \le \ l_{tc}, r_{tc} \ & \le \ \maxt && \forall c\in [p], t\in [d], \label{eq:cp:dom} \\
    z_{ic} \ & \in \ \{ 0, 1 \} && \forall i\in [n], c\in [p]. \label{eq:cp:end}
\end{align}
Again, the objective function \eqref{eq:cp:begin} asks the total span to be minimized. For $i\in [n]$, $c\in [p]$, and $t\in [d]$, constraints \eqref{eq:cp:constr1}-\eqref{eq:cp:constr2} assert that the cluster bounds  include the point $x^i$ in every coordinate if the point $x^i$ is assigned to the cluster $c$. Constraints~\eqref{eq:cp:exactlyone} ask every point to belong to exactly one cluster. In this expression we use the construct \textsc{ExactlyOne}, which is usual in constraint programming environments, and asserts that exactly one variable from its arguments must take a nonzero value. Finally, constraints \eqref{eq:cp:dom}-\eqref{eq:cp:end} specify the domains for the variables.

\subsection{Related work}

The hyper-rectangular clustering problem has been addressed in the
literature using several exact optimization approaches. Early formulations are
based on mixed-integer programming models that explicitly assign each point to a
cluster and determine the corresponding rectangular bounds (\cite{mago12,park09}). These formulations
provide a natural and flexible modeling framework, but they suffer from strong
symmetry issues and quickly become computationally challenging as the number of
points increases.

To improve scalability, further exact methods have been proposed. In
particular, branch-and-cut approaches exploit valid inequalities to strengthen the
linear relaxation (\cite{marenco23}), while extended formulations combined with branch-and-price
techniques allow the consideration of exponentially many variables in a dynamic
fashion (\cite{clustering_bap}). These methods significantly improve performance compared to compact
formulations and represent the current state of the art for solving HRCP to
optimality.

Despite these advances, all existing exact approaches share a common limitation:
they fundamentally rely on solving the problem over the full dataset. As a
consequence, their computational performance remains strongly dependent on the
number of points, which limits their applicability to large-scale instances (\cite{marenco23,clustering_bap}).

The approach proposed in this paper departs from this paradigm. Instead of solving
HRCP on the full set of points, we exploit a structural property of the problem:
optimal cluster boundaries are determined by a subset of points, while interior
points do not influence the objective value. This allows us to reformulate the
solution process as a sequence of smaller subproblems defined on adaptively
selected subsets of points.

This idea is reminiscent of approaches that exploit reduced representations of the data,
such as core-set constructions in clustering (\cite{badoiu2002approximate,agarwal2005geometric, feldman2011unified}).
These methods aim at identifying a small subset of representative points that approximately
preserve the structure of the original dataset. However, unlike these approaches, which
typically provide approximation guarantees, our method preserves exact optimality: once
the selected subset satisfies a simple coverage condition, the solution is guaranteed to
be optimal for the full dataset.

Our approach is also related, at a conceptual level, to decomposition techniques
such as column generation and constraint generation in mathematical programming.
These methods iteratively refine a formulation by dynamically adding variables or
constraints that are relevant for optimality. In contrast, our method does not
modify the formulation itself, but rather reduces the effective size of the input
by selecting a subset of points on which the problem is solved. In this sense, the
incremental procedure can be seen as identifying the data points that are necessary
to define the optimal solution, instead of the variables or constraints that are
necessary to describe it.

% The perspective introduced in this paper is fundamentally different from previous exact methods. While branch-and-cut and branch-and-price techniques aim at improving the solution of a fixed formulation, our approach reduces the effective size of the instance by identifying the points that are relevant for defining the optimal solution.

\section{The incremental algorithm}\label{sec:algorithm}

In this section we provide the main methodological proposal of this work, by presenting an adaptive exact strategy based on the ideas outlined previously. This method tries to take advantage of the capacity to solve small instances to optimality of previous approaches, and incrementally solves sub-instances of the original instance while trying to achieve optimality. Subsection~\ref{sec:components} summarizes the main components of the proposed procedure. Subsection~\ref{sec:repairing} presents an optional repairing heuristic to enhance the search for feasible solutions. Finally, Subsection~\ref{sec:samplingmetrics} reviews potential sampling metrics for obtaining sub-instances of the instance to be solved.

\subsection{The incremental procedure}
\label{sec:components}

Our algorithm starts by solving an instance with a small subset of points and iteratively adds more points if these are not covered by the obtained solution. We prove that as soon as a solution covers the whole set of point from the instance, then the solution is indeed an optimal solution for the original problem.
Algorithm \ref{alg.framework} illustrates the general framework of the proposed methodology.
Given a $p$-clustering $\mathbb{C}$ of a subset of points $\hat \X \subseteq \X$, we say that $\mathbb{C}$ \emph{covers} $\X$ if every point $x \in \X$ lies inside at least one of the hyper-rectangles induced by $\mathbb{C}$ (i.e., those enclosing each set of points of the partition). With this definition, the following result allows us to prove the correctness of Algorithm 1.

\begin{algorithm}
    \caption{General framework for the incremental algorithm}\label{alg.framework}
    \begin{algorithmic}[1]
    % {\linespread{1}\selectfont
			\State Choose an initial subset of points $\hat \X \subseteq \X$ \label{step:initialize}
			
			\State Solve the clustering on $\hat \X$ and get an optimal solution $\mathbb{C}$ \label{step:solve}

			\If{$\mathbb{C}$ covers $\X$} \label{step:check_feasible}
			    \State Add each point $x \in \X \setminus \hat \X$ to the cluster from $\mathbb{C}$ enclosing $x$
				\State \Return $\mathbb{C}$ \label{step:return}
			\Else
				\State Add to $\hat \X$ some points from $\X \setminus \hat \X$ \label{step:addpoints}
				\State \textbf{go to} Line \ref{step:solve}
			\EndIf
    % }
    \end{algorithmic}
\end{algorithm}

\begin{lemma} \label{lem:opt}
Let $\hat \X \subseteq \X$ be a subset of points and $\mathbb{C}$ an optimal $p$-clustering of $\hat \X$ (i.e., with minimum total span). If $\mathbb{C}$ covers $\X$, then it represents an optimal $p$-clustering for $\X$.
\end{lemma}

\noindent
The intuition behind this result is that clusters are determined by the points defining the boundaries of their hyper-rectangles.
Therefore, if these extreme points are already present in the subset $\hat \X$, adding interior points
cannot modify the optimal rectangles and thus cannot improve the objective value.

\begin{proof}
Since $\mathbb{C}$ covers $\X$, we can add each point of $\X \setminus \hat \X$ to some cluster from $\mathbb{C}$, thus yielding a feasible $p$-clustering for $\X$ without changing the hyper-rectangles induced by the clustering (i.e., with a total span equal to $\mbox{span}(\mathbb{C})$).
Assume there exists another $p$-clustering $\mathbb{C}'$ of $\X$ with $\mbox{span}(\mathbb{C}') < \mbox{span}(\mathbb{C})$. Since $\mathbb{C}'$ is a $p$-clustering of $\X$, it also covers $\hat \X$, thus contradicting the fact of $\mathbb{C}$ being optimal for $\hat X$.
\end{proof}

An alternative proof for Lemma \ref{lem:opt} follows from the fact that \prob{} over $\hat \X$ is indeed a combinatorial relaxation of \prob{} over $\X$, as it allows points from $\X \setminus \hat \X$ to be left uncovered by the solution. 
More precisely, it is trivial to see that every feasible solution for \prob{} over $\X$ gives also a feasible solution for \prob{} over $\hat \X$.
Hence, if an optimal solution for $\hat \X$ (i.e., the relaxation) is also feasible for $\X$ (i.e., the original problem), then this solution is indeed optimal for $\X$. 
Remark \ref{rem:lb} follows from this fact.

\begin{remark} \label{rem:lb}
Any lower bound for \prob{} over $\hat \X \subseteq \X$ 
% (e.g., obtained in Step \ref{step:solve} of Algorithm \ref{alg.framework}) 
is a valid lower bound for \prob{} over $\X$.
\end{remark}

\begin{proposition} 
Algorithm \ref{alg.framework} always finds an optimal solution for \prob{} over $\X$.
\end{proposition}

\begin{proof}
The algorithm finishes only when the condition on Line \ref{step:check_feasible} is satisfied, i.e., when $\mathbb C$ covers $\X$. By Lemma \ref{lem:opt}, this provides an optimal solution for \prob{} over $\X$, since $\mathbb{C}$ is an optimal $p$-clustering of $\hat \X \subseteq \X$. So it just remains to prove that the algorithm reaches this point at least once. 

The set $\hat \X$ is increased on every iteration by adding to it points from $\X \setminus \hat \X$, hence the algorithm will iterate until $\hat \X = \X$, if it does not finish earlier. At this point, the clustering $\mathbb{C}$ obtained in Line \ref{step:solve} does trivially cover $\X$, hence it is returned (as it is optimal) in Line \ref{step:return}.
\end{proof}

A strong characteristic of Algorithm~\ref{alg.framework} is the fact that as soon as a feasible solution for $\X$ is found (i.e., in Step~\ref{step:check_feasible}), the algorithm stops as this solution is optimal. 
On the other hand, if the process is stopped before its termination (e.g., due to an imposed time limit), no feasible solution (even if sub-optimal) can be returned.
To tackle this issue, we may profit from the several executions of Step \ref{step:solve} which tries to find optimal solutions for \prob{} over $\hat \X$. During this process, the procedure may generate intermediate solutions ($p$-clusterings) that are feasible for $\hat \X$ (for example, sub-optimal incumbent solutions during the branch-and-bound process when solving Step~\ref{step:solve} using a MIP solver). Even if the optimal solution on this step does not provide a cover for $\X$, these intermediate solutions may do.
Therefore, we can inspect these solutions searching for feasible solutions for \prob{} over $\X$, thus keeping track of an incumbent (i.e., the current best feasible solution found) in Algorithm \ref{alg.framework}. 

\begin{remark} \label{rem:ub}
    Any \textbf{feasible} solution for $\hat\X$ found during Step \ref{step:solve} of Algorithm \ref{alg.framework} may represent a feasible solution for the original instance $\X$ and may help to improve the incumbent, which in turn gives an upper bound for the problem. 
    % This incumbent can be used as a warm start for Step \ref{step:solve}. 
\end{remark}
\noindent

Additionally, in the case in which a solution $\mathbb C$ for $\hat\X$ found during Step \ref{step:solve} of Algorithm \ref{alg.framework} is not a feasible solution for the original instance $\X$, it is always possible to extend $\mathbb C$ to a feasible solution for $\X$ by adding uncovered points to arbitrary clusters from $\mathbb C$. This goal may be achieved in several ways, and we propose this as an optional element of our algorithm. We describe a simple heuristic for this step in Section~\ref{sec:repairing}. \\

By Remark \ref{rem:lb}, any lower bound for \prob{} over $\hat \X$ obtained on Step \ref{step:solve} can be used as a lower bound for \prob{} over $\X$. In addition, by using the upper bound described in Remark \ref{rem:ub}, Algorithm \ref{alg.framework} can keep track of the optimality gap at any time. This gap can be closed even if the solution $\mathbb C$ found in Step \ref{step:solve} is not feasible for $\X$ (i.e., the incumbent from previous iterations may be proved to be optimal even if it does not match with $\mathbb C$). 
These additions can be included in the algorithm, thus generating the procedure described in Algorithm \ref{alg.withbounds}.

\begin{algorithm}
    \caption{The complete incremental algorithm}\label{alg.withbounds}
    \begin{algorithmic}[1]
    % {\linespread{1}\selectfont
		\State {Choose a an initial subset of points $\hat \X \subseteq \X$} \label{alg2.initialize}
		\State $LB$ $\gets$ 0, $UB$ $\gets$ $+\infty$, \emph{incumbent} $\gets$ null, \emph{tlim} $\gets\tau_0$
		
		\While{$LB < UB$}
		
			\State {Solve HRCP on $\hat \X$ (potentially with a time limit) and get solution $\mathbb{C}$} \label{alg2.solve}

			\If{$\mathbb{C}$ covers $\X$ and is optimal for $\hat \X$} \label{alg2.check_feasible}
			    \State Add each point $x \in \X \setminus \hat \X$ to a cluster from $\mathbb{C}$ enclosing $x$
				\State \Return $\mathbb{C}$ \label{alg2.return}
            \ElsIf{$\mathbb{C}$ covers $\X$ but the time limit (\emph{tlim}) was met}
                \State Increase \emph{tlim} \label{alg2.increasetlim}
			\Else \Comment{$\mathbb{C}$ does not cover $\X$} \label{alg2.nocover}

				\State $LB \gets \max\{LB,$ lower bound from {Line \ref{alg2.solve}}$\}$
				
				\For{each feasible solution $\mathbb{C}'$ found at {Line \ref{alg2.solve}}}
                    \If{$\mathbb{C}'$ does not cover $\X$}
					\State (Optional) \emph{Repair} $\mathbb{C}'$ to make it cover $\mathcal X$ \label{alg2.repairing}
				    \EndIf
				    \If{$\mathbb{C}'$ covers $\X$ and is better than incumbent}
						\State Update $UB$ and \emph{incumbent} with $\mathbb{C}'$
					\EndIf
				\EndFor
			
				\State {Add to $\hat \X$ some points from $\X \setminus \hat \X$} \label{alg2.addpoints}
                \State \emph{tlim} $\gets\tau_0$ \label{alg2.resettlim}
			\EndIf
		\EndWhile
		\State \Return \emph{incumbent}
    % }
    \end{algorithmic}
\end{algorithm}

In Line~\ref{alg2.solve} of Algorithm~\ref{alg.withbounds} a time limit is imposed on the solution of HRCP for the sub-instance $\hat \X$ in order to keep running times in control. If the time limit is met, then the obtained solution $\mathbb{C}$ may not be optimal for $\hat \X$. Hence, even if $\mathbb{C}$ covers $\X$ we cannot apply Lemma~\ref{lem:opt} to claim that the clustering is optimal for $\X$. Additionally, since $\mathbb{C}$ covers $\X$ no candidate points can be added to $\hat \X$ in order to make progress by increasing this set. In this situation, the procedure increases the time limit (Line~\ref{alg2.increasetlim}) and retries the solution of HRCP for $\hat \X$ (Line~\ref{alg2.solve}). When $\mathbb{C}$ does not cover $\X$ (Line~\ref{alg2.nocover}), we increase $\hat \X$ with new points and reset the time limit to its original value $\tau_0$ (Line~\ref{alg2.resettlim}).

\subsection{A repairing heuristic to find feasible solutions}
\label{sec:repairing}

In Line \ref{alg2.repairing} of Algorithm \ref{alg.withbounds}, a repairing procedure is (optionally) applied to a solution $\mathbb C'$ for $\hat\X$ that does not cover $\X$. The goal is to obtain a feasible solution for $\X$, that may improve the incumbent. This problem is clearly NP-hard, as it generalizes \prob. 
Since this procedure is applied multiple times throughout the algorithm, we propose to use a simple heuristic, the running of which would have a minimal impact on the overall process.

The heuristic aims to iteratively add uncovered points to the clustering $\mathbb{C}$ by greedily selecting the uncovered point closest to the clustering, i.e., the point that would need the smallest enlargement of a cluster from $\mathbb{C}$ for being included in the clustering. Formally, we define the \emph{distance of a point $x$ to the clustering $\mathbb{C}$} as:
$$
\mathrm{dist}(x, \mathbb{C}) = \min_{C \in \mathbb{C}} \left\{ \sum_{t=1}^d \mathrm{dist}_t(x, C)\right\},
$$
where 
% $\mathrm{dist}_t(x, C) = \max(x_t - \max_t(C), \min_t(C) - x_t, 0)$
$\mathrm{dist}_t(x, C) := \max\{x_t - \max\{x'_t : x' \in C\}, 0, \min\{x'_t : x' \in C\} - x_t\}$
represents the distance of the point $x$ to the cluster $C$ over the coordinate $t$.
The resulting heuristic is depicted in Algorithm \ref{alg.repairing}.

\begin{algorithm}
    \caption{Repairing heuristic}\label{alg.repairing}
    \begin{algorithmic}[1]
    % \color{blue}
    % {\linespread{1}\selectfont
		\State {\textbf{Input:} A clustering $\mathbb{C}$ covering a subset of points $\hat \X \subseteq \X$}\\
		\State $\mathcal X'$ $\gets$ subset of uncovered points $\X \setminus \hat \X$
		\While{$\mathcal{X}' \neq \emptyset$}
		
			\State $\hat x \gets \arg \min_{x \in \mathcal X'} \{ \mathrm{dist}(x, \mathbb{C}) \}$, i.e., the point in $\mathcal{X}'$ closest to $\mathbb{C}$

                \State Add $\hat x$ to the closest cluster from $\mathbb{C}$, thus expanding the cluster 

                \State Remove $\hat x$ from $\mathcal{X}'$
		\EndWhile
		\State \Return incumbent
    % }
    \end{algorithmic}
\end{algorithm}

\subsection{Sampling metrics}\label{sec:metrics}
\label{sec:samplingmetrics}

Lines \ref{alg2.initialize} and \ref{alg2.addpoints} from Algorithm \ref{alg.withbounds} add new points to $\hat \X$, by performing a \emph{sampling} of the points in $\X \backslash \hat \X$.
We refer to these lines as the \emph{initialization} and \emph{expansion} steps, respectively. These steps can be performed in many different ways. 

\begin{figure}
    \centering
    \includegraphics[width=.45\textwidth]{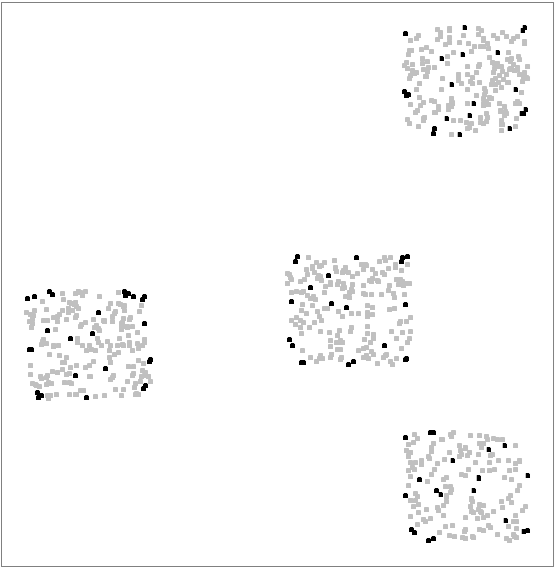}
    \hfill
    \includegraphics[width=.45\textwidth]{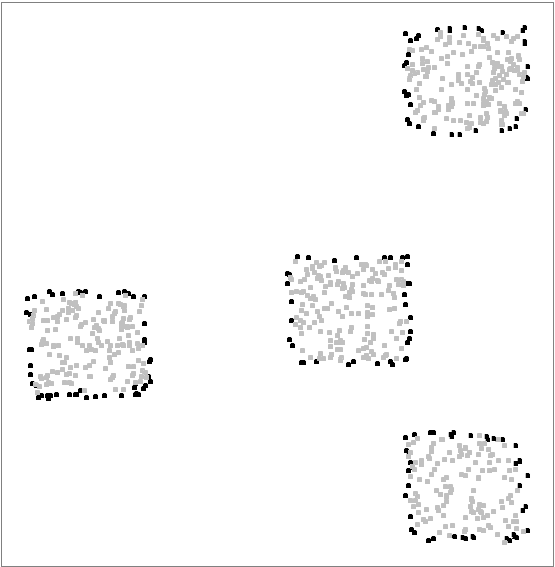}
    \caption{A small example in $\real^2$ illustrating the difference between a uniform random sample (left) and a sample which identifies points which are most likely to lie in the borders of the hyper-rectangles defining the optimal solution (right). Black points represent the sample $\hat \X$ and gray points are the remaining points $\X \setminus \hat \X$.}
    \label{fig:sampling}
\end{figure}

The most straightforward sampling method is a \emph{random sampling}. For the initialization step, we may set $\bar \X$ to a random sample, e.g., by uniformly choosing points from $\X$ with some fixed probability. 
The same can be done for the expansion step, choosing points from $\X \setminus \X$ which are not covered by $\mathbb C$.  
Unfortunately, although not unexpected, preliminary experimentation proved this method to be inefficient in practice. 
A random sampling tends to conserve the spatial distribution of the data points, however, the best sampling method should be able to identify the points which are most likely to lie in the perimeter of one of the hyper-rectangle which encloses a cluster in the solution, as these are the points ``defining'' the solution. 
Figure \ref{fig:sampling} shows a small example in $\real^2$ illustrating the difference between a uniform random sample (left) and a sample which identifies points which are most likely to lie in the borders of the hyper-rectangles defining the optimal solution (right). Black points represent the sample $\hat \X$ and gray points are the remaining points $\X \setminus \hat \X$.

In this section we explore different sampling approaches and propose different alternatives to perform these sampling steps, thus obtaining different versions of the algorithm. 
Throughout the remainder of this work, let $N: \X \rightarrow 2^{\X}$ be any \emph{neighbourhood function} on $\X$. In particular, we will use $N(x) = \{ y \in \X \setminus \{x\} : \|x - y\|_2 \leq \delta\}$, for a fixed value of $\delta \in \real_+$, however, every method proposed in this paper can be applied for other neighbourhood functions.

\subsubsection{Neighbourhood metric}\label{sec:neighmetric}

Identifying points in the border of a cluster is similar to the classical DBSCAN algorithm~(\cite{dbscan}) for density-based clustering, so we use its notion of \emph{core} and \emph{border} points. In this classical method, a value $minPts \in \mathbb N$ is chosen, and a point $x \in \X$ having $|N(x)| \geq minPts$ is called a \emph{core} point, and it is considered to have high chances of lying completely in the interior of the cluster which contains it. In other words, this point is very likely to not lie in the border of the cluster. All points that are not core points are defined to be \emph{border} points.\footnote{More precisely, in DBSCAN a \emph{border} point has $0<|N(x)|<minPts$, and if $|N(x)|=0$ then is called an \emph{outlier} point. Since we are not allowing outliers in our problem, we omit this concept for clarity.}
The algorithm proposed in the referenced work applies a heuristic which collects core points in a smart way in order to detect the clusters in the solution.

Even when the notions of border and core points may be useful in our setting, we further need something slightly more flexible than fixing a value $minPts$ to determine border points, as we need to refine this decision in the subsequent iterations of Algorithm \ref{alg.withbounds}.
Nevertheless, we observe that when points in a cluster are uniformly distributed in the space, the number of neighbours of a point $x \in \X$ is gradually reduced when the point is close to the border of the cluster, and is minimal for the points lying in the border.
Hence, we propose to use $|N(x)|$ as an (inverse) measure of how likely is a point to be in the border of its cluster. 
We denote this metric as the \emph{neighbourhood metric} and we use it as follows:
% \newcommand{\eps}{\varepsilon}
% \begin{definition}[\cite{dbscan}]
%     The $\eps$-neighborhood of a point $x \in \X$ is defined as 
%     $N_{\eps}(x) = \{y \in \X\ :\ \text{dist}(x,y) \leq \eps\}$. 
%     For some given $\eps \in \mathbb R_+$ and $\alpha \in \mathbb N$, point $x$ is a \emph{core point} if $|N_{\eps}(x)| \geq \alpha$. If $x$ is not a core point, then it is called a \emph{border point}.
% \end{definition}
\begin{itemize}
	\item \emph{Initialization step:}
	Let $x_{0} \in \X$ be a point with minimum value for $|N(x_0)|$. 
	Set $\hat X = \{ x \in \X : |N(x)| \leq \alpha |N(x_0)|\}$, for a given $\alpha \geq 1$.

	\item  \emph{Increment step:}
	Sort points $\X \setminus \hat \X$ which are not covered by the obtained solution $\mathbb C$, and choose those with fewer neighbours, up to a limit of $k\in \mathbb N$ points.
\end{itemize}

\subsubsection{Eccentricity metric}\label{sec:eccmetrics}

The neighbourhood metric proposed in Section \ref{sec:neighmetric} may not be very accurate identifying border points when different clusters have different densities; a border point from a very dense cluster may have more neighbours than a core point from a cluster with low density. Moreover, most of the new points added to $\hat\X$ will be part of dense clusters, hence providing a biased information. To tackle this issue, we propose a metric that considers the position of a point relative to the position of its neighbours, taking into account that border points usually have every neighbour towards the center of the cluster.   

For a given coordinate $t \in [d]$, we define the \emph{lower} and \emph{upper neighbourhoods} of a point $x \in \X$ in dimension $t$ as $N^-_t(x) = \{ y \in N(x) : y_t \leq x_t \}$ and $N^+_t(x) = \{ y \in N(x) : y_t > x_t \}$, respectively.

\begin{definition}
The \emph{eccentricity} of a point $x \in \X$ (with $|N(x)|>0$) in dimension $t \in [d]$ is
\begin{equation}
    E^t(x) = \frac{\max\{ |N^-_t(x)|, |N^+_t(x)| \}}{|N(x)|}
\end{equation}
and its global \emph{eccentricity} is $E(x) = \max_{t \in [d]} \{E^t(x)\}$. If $|N(x)|=0$ then $E^t(x)=1$.
\end{definition}
\begin{figure}
    \centering
    \begin{minipage}{0.32\textwidth}
    		\begin{center}
    			{\small \includegraphics[width=\textwidth]{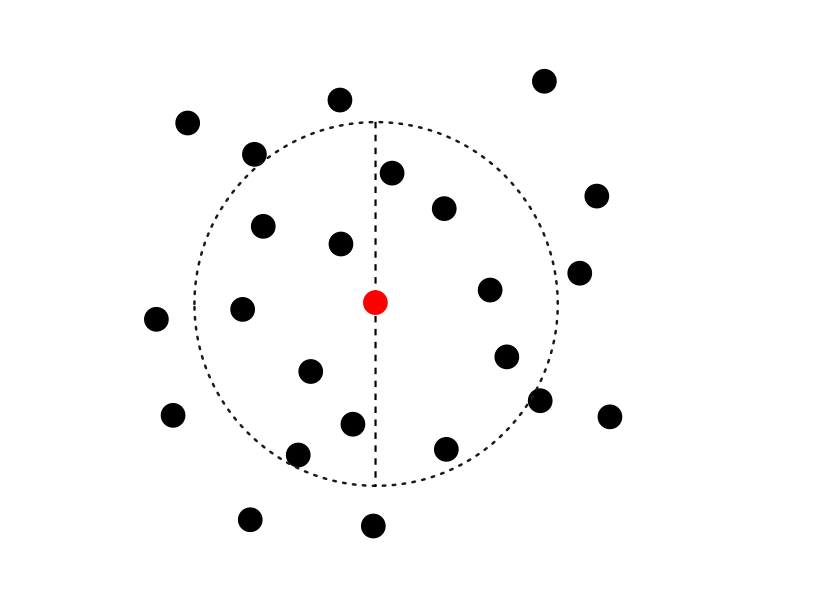}\\
    			$E^1(x) = \frac{6}{11} \approx 0.55$}%
    		\end{center}        
    \end{minipage}
    \vrule
    \begin{minipage}{0.32\textwidth}
    		\begin{center}
    			{\small \includegraphics[width=\textwidth]{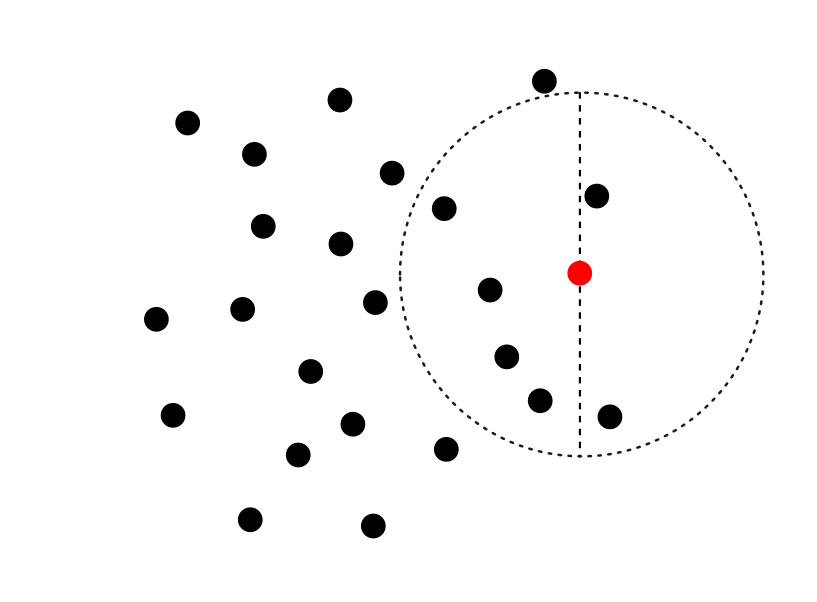}\\
    			$E^1(x) = \frac{4}{6} \approx 0.67$}%
    		\end{center}        
    \end{minipage}
    \vrule
    \begin{minipage}{0.32\textwidth}
    		\begin{center}
    			{\small \includegraphics[width=\textwidth]{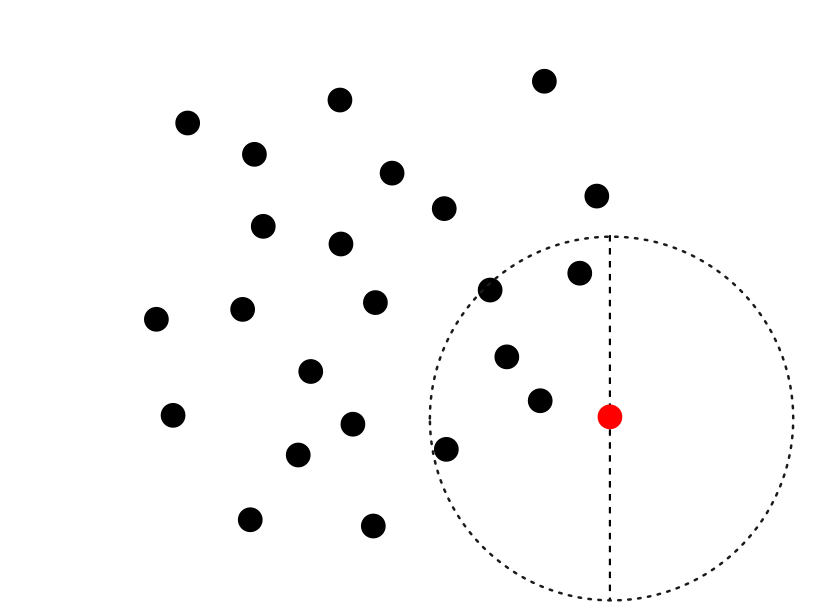}\\
    			$E^1(x) = \frac{5}{5} = 1$}%
    		\end{center}        
    \end{minipage}

    \caption{Eccentricity (fixed on the horizontal coordinate, $t=1$ in the example) of different points in a cluster as the points are taken closer to the border of the cluster.}
    \label{fig:eccentricity}
\end{figure}

\noindent
Figure \ref{fig:eccentricity} depicts the eccentricity (fixed on the horizontal coordinate, $t=1$ in the example) of different points in a cluster as the points are taken closer to the border of the cluster. The point chosen on the first case (left) is a very centric point and has an eccentricity of $E^1(x) = \frac{6}{11} \approx 0.55$. The second example (middle) is closer to the border and its eccentricity is $E^1(x) = \frac{4}{6} \approx 0.67$. Finally, the third example (right) lies in the border of the cluster, and its eccentricity is $E^1(x) = \frac{5}{5} = 1$.
The eccentricity of any point is a value between 0.5 and 1, regardless of the density of the clusters and the chosen neighbourhood function $N$. 
The \emph{eccentricity metric} for Algorithm \ref{alg.withbounds} works as follows:
\begin{itemize}
	\item \emph{Initialization step:}
	Let $x_{0} \in \X$ be a point with maximum global eccentricity $E(x_0)$. 
	Set $\hat X = \{ x \in \X : E(x) \geq \beta E(x_0)\}$, for a given $\beta \in [0,1]$.

	\item  \emph{Increment step:}
	Sort points $\X \setminus \hat \X$ which are not covered by the obtained solution $\mathbb C$, and choose those with higher global eccentricity, up to a limit of $k\in \mathbb N$ points.
\end{itemize}

\subsubsection{Distance-Eccentricity metrics}\label{sec:disteccmetrics}

\begin{figure}   
    \centering
    \hfill
    \begin{minipage}{0.35\textwidth}
    		\begin{center}
    			\includegraphics[width=.7\textwidth]{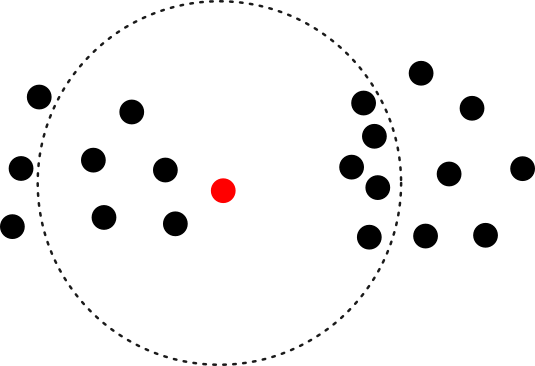}
    		\end{center}        
    \end{minipage}
    \hfill
    \begin{minipage}{0.5\textwidth}
    		\begin{center}
                $|N(x)|$ is high $\Rightarrow$ neighbour metric fails \\
				$E(x)$ is low $\Rightarrow$ eccentricity metric fails 
    		\end{center}        
    \end{minipage}
    \hfill

    \caption{Example in which a point which lies in the border of a cluster is not detected either by the neighbour metric nor by the eccentricity metric.}
    \label{fig:failure}
\end{figure}

Although the eccentricity metric might properly deal with issues caused by the presence of clusters with different densities, both this and the neighbourhood metrics suffer from another issue, related to the distance between clusters.
Figure \ref{fig:failure} shows an example in which a point in the border of a cluster is not detected either by the neighbour metric nor by the eccentricity metric. This example shows the importance of the definition of $N(x)$ and its impact on these metrics (e.g., a bad choice of the parameter $\delta$ defining the size of the neighbourhood when clusters are not much separated). 

With the goal of dealing with this type of situations, we propose another metric that considers not only the number of neighbours on ``each side'' but also the average distance to them. More precisely, a candidate point for being a border point is a point in which the distance to the points in one side of its neighbourhood is very different to the distance to the other side. We formalize this concept in the following definition.

\begin{definition}
The \emph{distance-eccentricity} of a point $x \in \X$ in dimension $t \in [d]$ is
\begin{equation}
    D^t(x) = |\text{\rm avgdist}_t(x, N^-_t(x)) - \text{\rm avgdist}_t(x, N^+_t(x))|,
\end{equation}
where $\text{\rm avgdist}_t(x, X) = \frac{1}{|X|}\sum_{y \in X} |y_t - x_t|$ is the average distance between $x$ and $X\subseteq \X$ on coordinate $t\in [d]$.
The global \emph{distance-eccentricity} is $D(x) = \sum_{t \in [d]} \{D^t(x)\}$.
\end{definition}

The \emph{distance-eccentricity metric} for Algorithm \ref{alg.withbounds} works analogously to the eccentricity metric:
\begin{itemize}
	\item \emph{Initialization step:}
	Let $x_{0} \in \X$ be a point with maximum global distance-eccentricity $D(x_0)$. 
	Set $\hat X = \{ x \in \X : D(x) \geq \beta D(x_0)\}$, for a given $\beta \in [0,1]$.

	\item  \emph{Increment step:}
	Sort points $\X \setminus \hat \X$ which are not covered by the obtained solution $\mathbb C$, and choose those with higher global distance-eccentricity, up to a limit of $k\in \mathbb N$ points.
\end{itemize}
\bigskip

We conclude this section noting that none of the metrics introduced in this work depends on the current elements in the set $\hat \X$. Therefore, the values required by these metrics can be computed at the beginning of the algorithm with a negligible impact on its performance.

\section{Computational Experiments}\label{sec:experimentation}

We report in this section our computational experience with the incremental algorithm presented in Section~\ref{sec:algorithm} and the sampling metrics introduced in Subsection~\ref{sec:metrics}. The goal of the experiments is twofold: first, to evaluate the scalability of the incremental framework with respect to the number of points, and second, to assess the impact of the sampling strategy used to construct the subsets of points solved during the algorithm.

For the solution of \prob{} over a subset of points $\hat{\X}$ (Line~\ref{alg2.solve}
of Algorithm~\ref{alg.withbounds}), we use the compact formulation \eqref{cm.constr.0}--\eqref{cm.constr.7} proposed in \cite{marenco23} solved with a mixed-integer programming solver (\cite{cplex}). Alternatively, we solve the constraint programming model \eqref{eq:cp:begin}--\eqref{eq:cp:end} using the CP-SAT solver (\cite{cpsatlp}) from Google OR-Tools (\cite{ortools}). For benchmarking purposes we also consider a \emph{monolithic approach},
denoted as ``FULL'', in which the complete instance is solved directly without using the incremental framework.

The incremental procedure was implemented in Java and interfaced with Cplex
through the Concert API (\cite{cplex}). The source code of our implementation
is publicly available online\footnote{Available in the branch ``master'' at
\texttt{https://github.com/jmarenco/clusterswithoutliers}.}.

In order to conduct experiments with controlled instances having predictable optima, we use the instance generator introduced in \cite{marenco23}. This generator takes as input the dimension $d$, the number $n$ of points, the number $p$ of clusters, and a parameter $s\in[0,1]$ specifying the \emph{dispersion} of the clusters. First, a set of $p$ \emph{originating points} is generated uniformly at random in $[-1,1]^d$. These points act as the centers of the clusters. The remaining points are generated around these centers by sampling uniformly in hyper-rectangles of span $s$. More precisely, an originating point $\bar{x}$ is selected and a point is drawn uniformly from $\bar{x}+[-s/2,s/2]^d$.

Our experiments consider instances with $n\in\{400,1000,4000,10000\}$ points, dimensions $d\in\{2,3,5\}$, and number of clusters $p\in\{4,6\}$. For each parameter configuration we generate five instances using different random seeds, for a total of 120 instances.

Within the incremental framework we evaluate three sampling strategies for selecting points to be included in the subsets $\hat{\X}$: {ECC}, based on the eccentricity metric introduced introduced in Section~\ref{sec:eccmetrics}; {DIST}, based on the distance metric introduced in Section~\ref{sec:disteccmetrics}; and {RAND}, which selects points randomly as a baseline. The distance radius $\delta$ to define the neighbourhood of a point is set to $0.2$, $0.391$, and $0.719$ for $d=2$, $3$, and $5$, respectively. These values are selected to maintain the proportion between the volume of the neighbourhood $N(x)$ and the total volume $2^d$ for all dimensions $d$. Finally, the number of points $k$ to add in each iteration is set to 1\%, 2\%, 5\%, 10\%, and 20\% of the original number of points $n$. To simplify the presentation of the results, for each metric we only present the best result among these five values for $k$.

All runs are performed with a time limit of 1800 seconds, and we take an initial time limit of $\tau_0 = 300$ seconds for each run of the solver. This time limit is increased by 50\% each time the solution is not guaranteed to be optimal but covers the set $\hat \X$ (Line~\ref{alg2.increasetlim} in Algorithm~\ref{alg.withbounds}).

To simplify the presentation of the experimental results, each configuration
is denoted using labels of the form \texttt{METHOD/SOLVER/METRIC}.
The first component (\texttt{METHOD}) specifies the overall solution strategy:
\begin{itemize}
\item {FULL}: the monolithic approach;
\item {INC}: the incremental algorithm with no time limits for the subproblem solver (i.e., $\tau_0 = \infty$);
\item {INC5m}: the incremental algorithm with a five-minute initial time limit policy for the subproblem solver (i.e., $\tau_0 = 300$ seconds).
\end{itemize}
The second component (\texttt{SOLVER}) indicates which solver is used to tackle the optimization subproblems:
\begin{itemize}
	\item {SM}: the mixed-integer programming solver Cplex.
	\item {CP}: the CP-SAT solver from Google OR-Tools.
\end{itemize}
The third component (\texttt{METRIC}) specifies the sampling metric used within the incremental algorithm (ECC, DIST or RAND, as previously defined); this component is omitted for the monolithic approach.

For example, the label \texttt{INC/SM/ECC} denotes the incremental
algorithm using Cplex to solve each subproblem and the ECC metric to
select candidate points, whereas \texttt{FULL/CP} denotes the monolithic model solved directly with CP-SAT.

\subsubsection*{Running times for moderate-difficulty instances}

\begin{figure}
    \centering
    \includegraphics[width=0.9\textwidth]{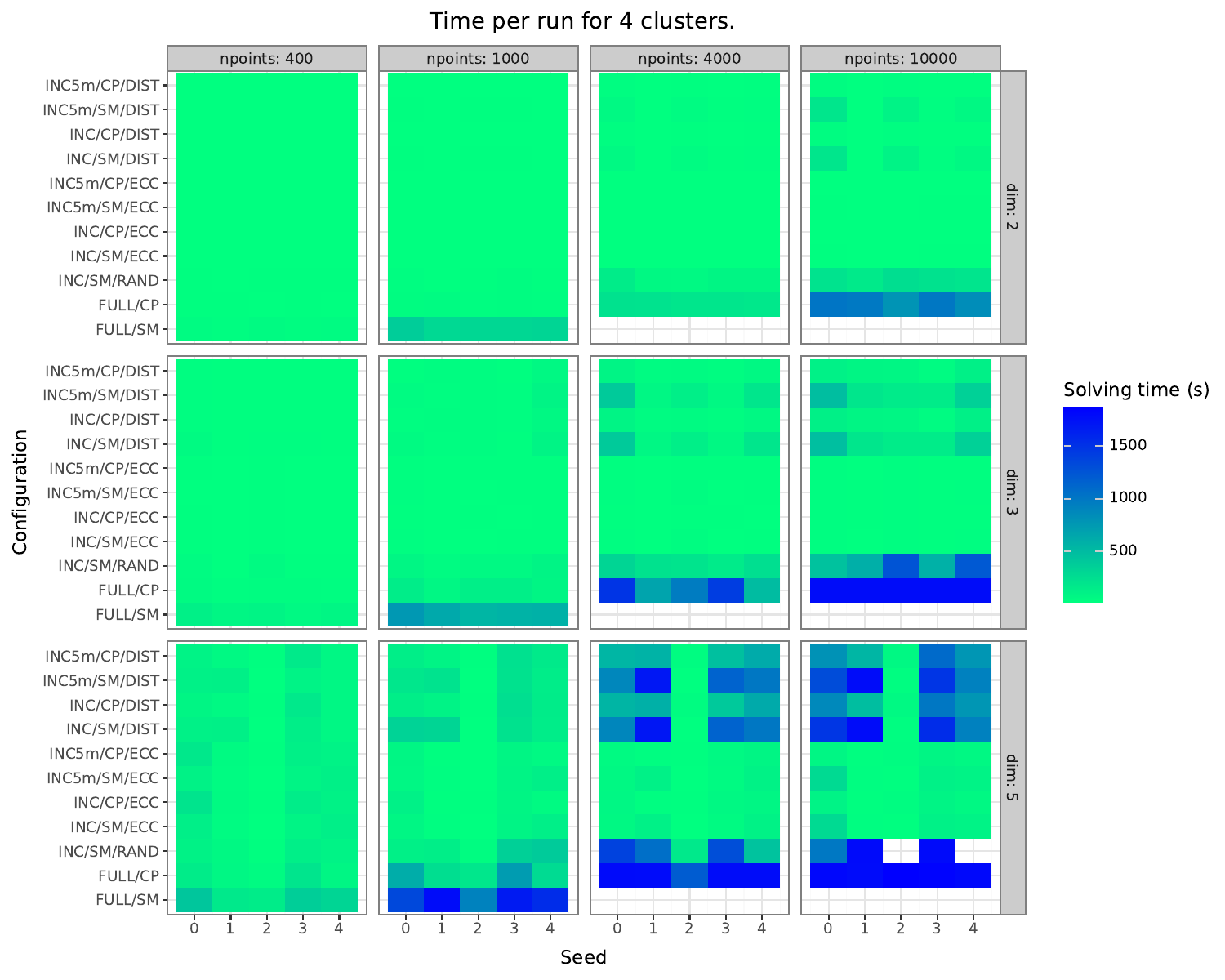}
    % \caption{Running times for instances with 4 clusters.}
    \caption{
        Running times for instances with $p=4$ clusters.
        Each cell corresponds to a single run of an algorithm on a specific instance.
        Columns represent instances with different numbers of points ($n \in \{400,1000,4000,10000\}$),
        while rows correspond to the tested algorithmic configurations.
        Colors represent the running time (in seconds, with a time limit of 1800) with darker colors indicating larger values. The missing cells correspond to runs in which no solution was found before the time limit.
        Each instance is tested with five different random seeds.
        The figure shows that the monolithic formulation (FULL) does not scale well as the number of
        points increases, while the incremental variants remain effective for much larger instances.
        Among the sampling strategies, the ECC metric consistently achieves the best performance,
        particularly on the largest instances. 
    }
    \label{fig:times4}
\end{figure}

Figure~\ref{fig:times4} reports the running times for instances with four clusters.
The results clearly show the limitations of the monolithic formulation as the instance size increases. While instances with a few hundred points can be solved within reasonable time, the compact MIP formulation quickly becomes intractable for larger instances. In particular, instances with more than 4000 points cannot be solved within the time limit using Cplex.

In contrast, the incremental approach scales significantly better with respect to the number of points. The algorithm remains able to solve instances with up to 10.000 points in many cases.

The experiments also highlight the importance of the sampling metric used to select candidate points. The variant based on random sampling (RAND) consistently performs worse than the metric-based strategies, indicating that identifying informative points is crucial for the efficiency of the incremental procedure.

Among the tested metrics, ECC provides the best performance, particularly for the largest and most challenging instances. This suggests that selecting points with high eccentricity is effective for identifying the points that define the boundaries of the clusters.

\subsubsection*{Increasing the number of clusters}

\begin{figure}[t!]
    \centering
    \includegraphics[width=0.9\textwidth]{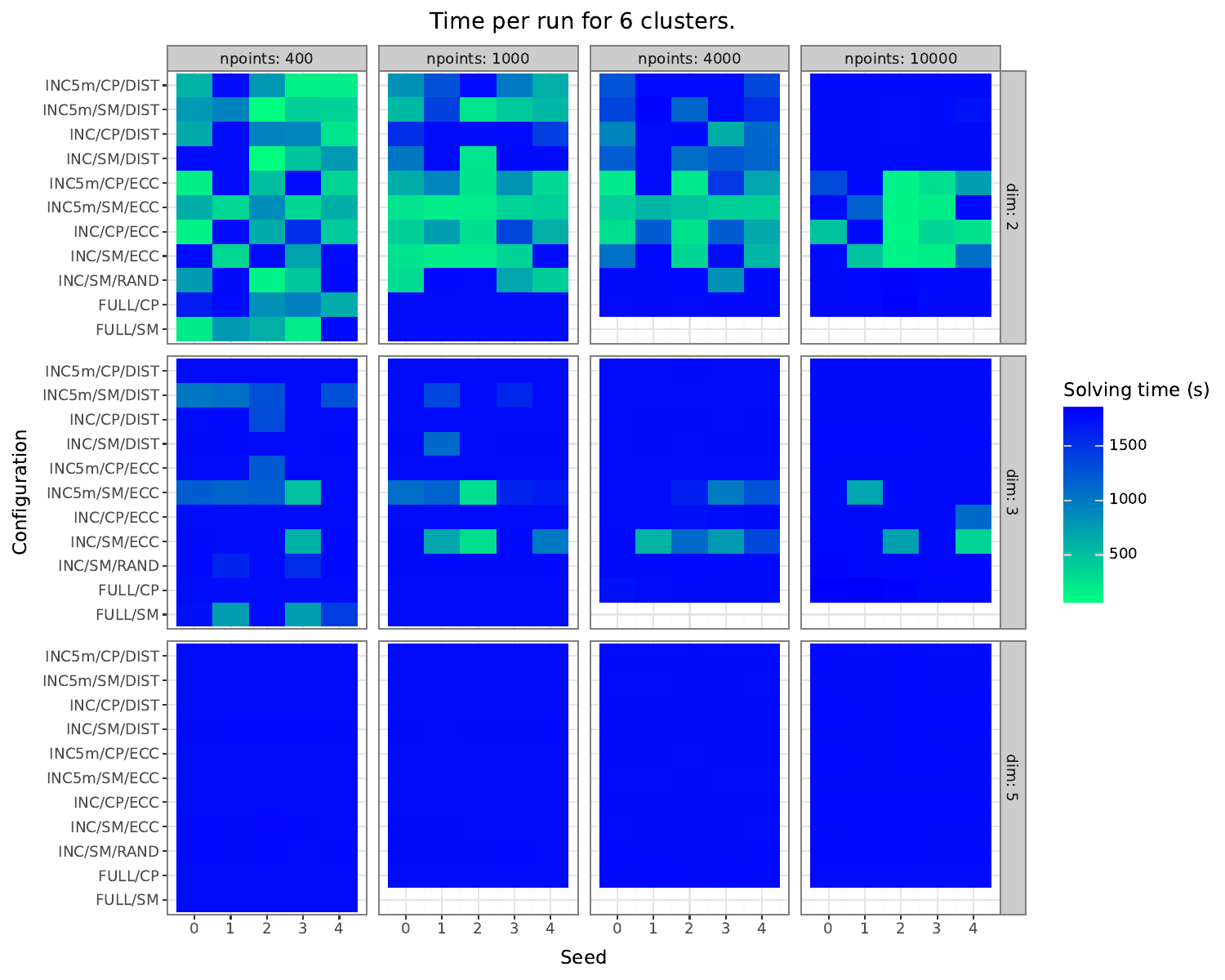}
    % \caption{Running times for instances with 6 clusters.}
    \caption{
        Running times for instances with $p=6$ clusters.
        Each cell represents a single run of an algorithm on a given instance.
        Columns correspond to instances with different numbers of points and rows to the tested
        algorithmic configurations.
        Colors indicate the running time in seconds.
        These instances are significantly harder than the $p=4$ case and only a small fraction of them
        can be solved to optimality within the time limit.
        The monolithic formulation fails to scale with the instance size, while the incremental
        variants remain more robust.
        As in the previous experiment, the ECC sampling strategy generally performs better than
	the alternatives.
    }
    \label{fig:times6}
    \end{figure}

We next consider a more challenging setting with six clusters, whose running times are shown in Figure~\ref{fig:times6}.
These instances are significantly harder. Only a small fraction of them can be solved to optimality within the time limit: some instances in dimension 2, very few in dimension 3, and none in dimension 5.

The monolithic formulation fails to solve almost all instances within the time limit, confirming once again the scalability limitations of this approach.
The incremental algorithms remain more robust, but the number of solved instances is still limited. As a consequence, running times alone are insufficient to compare the methods. In the following  we therefore analyze the quality of the solutions obtained by the algorithms. In this analysis we use two metrics for the quality of the solution: the \emph{reported gap} from each run, and \emph{real gap} computed using the best lower bound known for the problem. The first metric is probably more useful for a user interested in doing a single run, but with a reported quality guarantee. On the opposite, the second metric is more appropriate for a user only interested in the best solution, without requiring any quality guarantee for the solution.

\subsubsection*{Reported optimality gaps}

\begin{figure}
    \centering
    \includegraphics[width=0.9\textwidth]{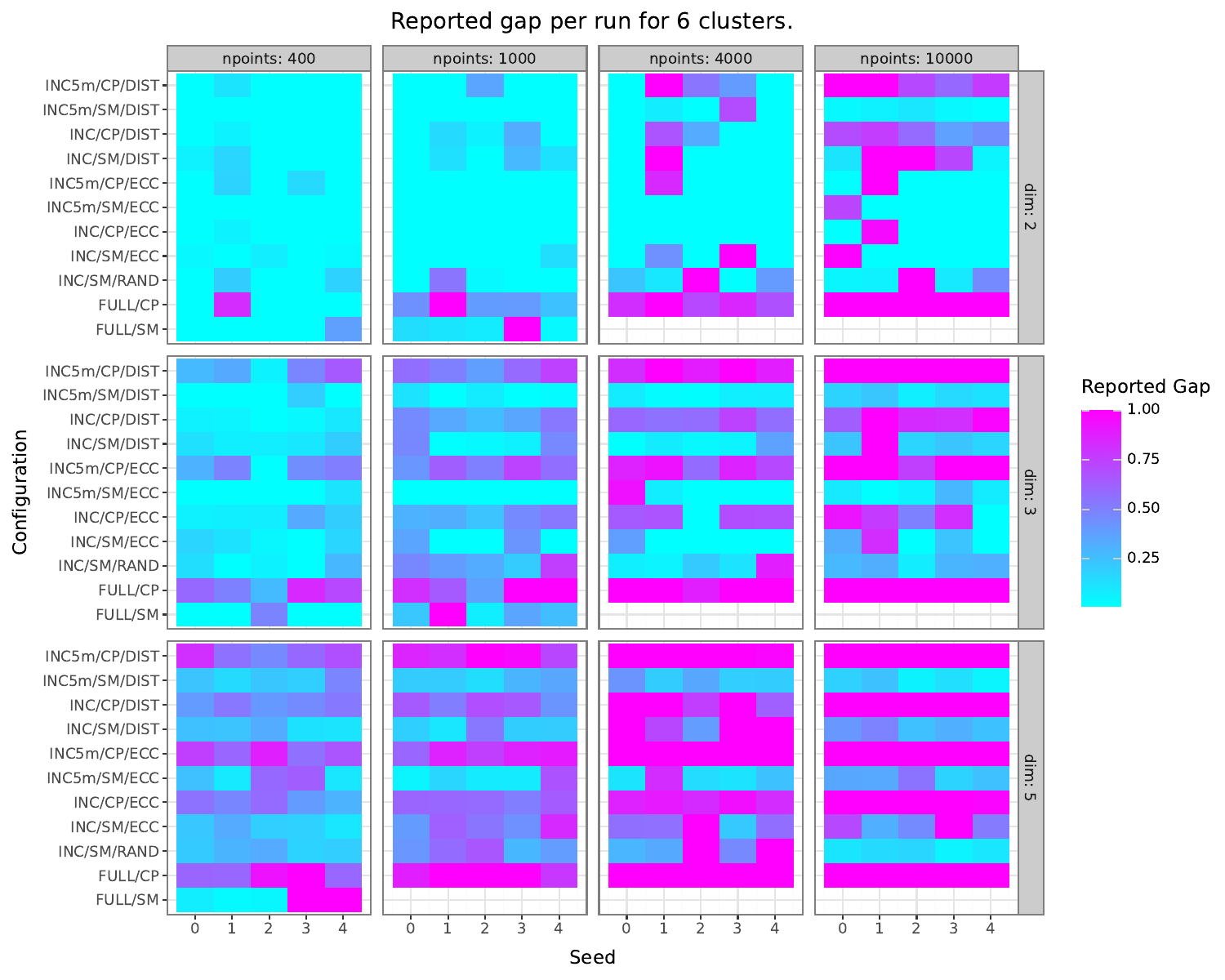}
    % \caption{Optimality gaps reported by each algorithm, computed using the lower bound produced internally by the solver used in each run.}
    \caption{
        Reported optimality gaps for instances with $p=6$ clusters.
        The gap is computed using the lower bound obtained internally by the solver in each run.
        Each cell corresponds to a single run of a given algorithm on one instance.
        Columns represent different instance sizes and rows correspond to the algorithmic configurations.
        Colors represent the magnitude of the reported gap.
        The results show that the problem becomes significantly harder as both the number of points
        and the dimension increase.
        Variants using the MIP solver (SM) typically report smaller gaps due to stronger dual bounds,
        while the ECC sampling strategy consistently outperforms the other sampling methods.
    }
    \label{fig:reportedgap}
\end{figure}

Figure~\ref{fig:reportedgap} presents the optimality gaps reported by each algorithm, computed using the lower bound produced internally by the solver used in each run.
The results reveal several interesting patterns.

First, the difficulty of the problem increases rapidly with both the dimension and the number of points. In particular, instances with dimension $5$ remain challenging even for relatively small values of $n$.

Second, the variants using the MIP solver (SM) typically report smaller optimality gaps than those using CP-SAT. This behavior is expected, as MIP solvers generally provide stronger dual bounds, while CP-SAT is designed primarily to find high-quality feasible solutions.

Third, the choice of sampling metric has a clear impact on performance. The ECC strategy consistently achieves smaller gaps than the alternatives, while the RAND variant performs noticeably worse. This confirms that the sampling strategy plays a critical role in the efficiency of the incremental framework.

Finally, introducing the internal time limit (the INC5m variants) often improves the results, suggesting that restarting the solver helps to explore different regions of the solution space. For instances in dimensions $2$ and $3$, the best configuration runs end with an optimal solution on 100\% and 85\% of the instances, respectively.

\subsubsection*{Real optimality gaps}

\begin{figure}
    \centering
    \includegraphics[width=0.9\textwidth]{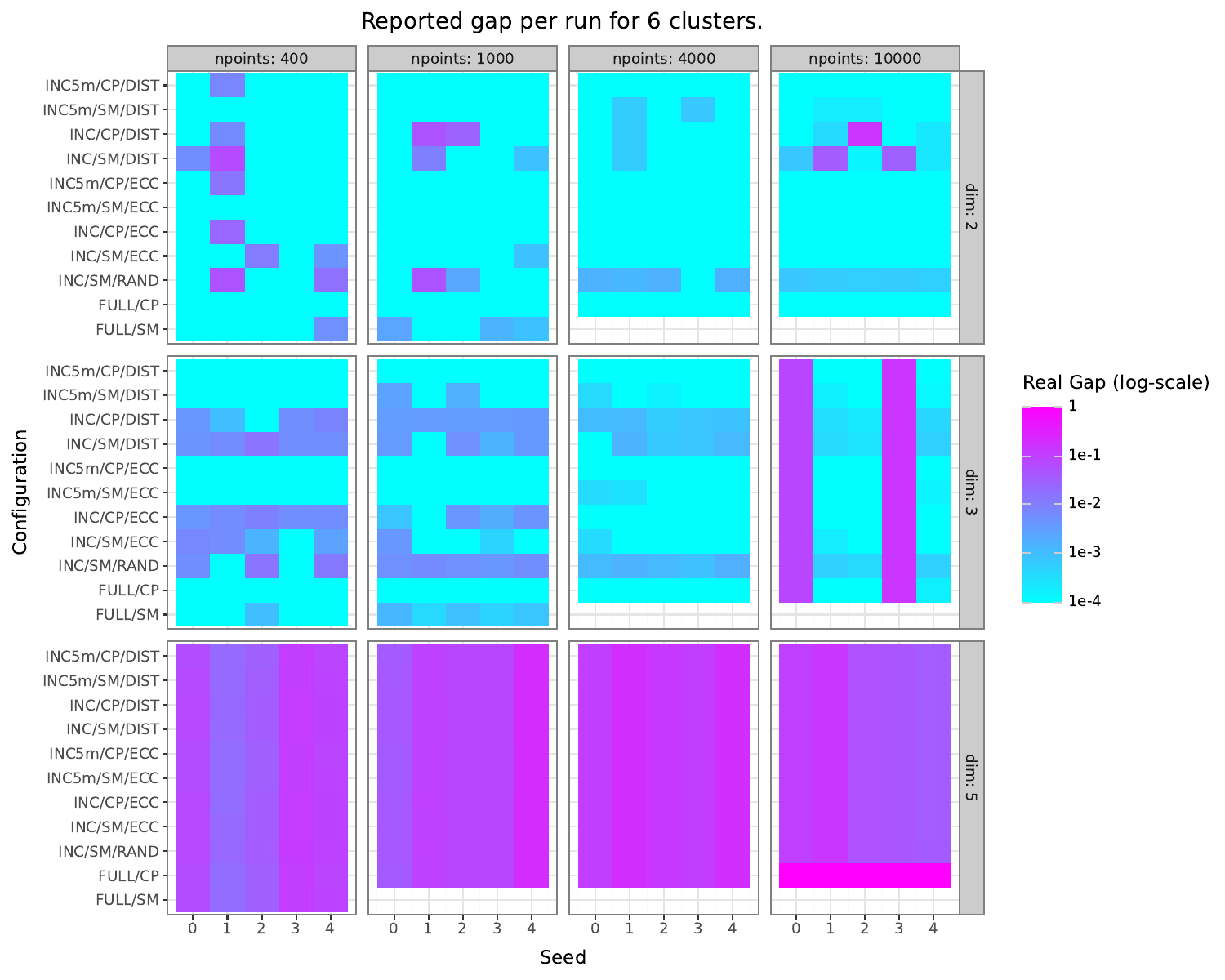}
    % \caption{Real gaps (colors in logarithmic scale) achieved by each algorithm, computed using the best lower bound obtained across all runs and configurations.}
    \caption{
        Real optimality gaps for instances with $p=6$ clusters.
        The gap is computed using the best lower bound obtained across all algorithms and runs.
        A logarithmic color scale is used to highlight differences between solutions.
        Although some algorithms report relatively large gaps due to weak lower bounds,
        the real gaps show that many solutions are in fact very close to optimal.
        The ECC strategy again provides the most reliable performance, while the random sampling
        variant (RAND) performs noticeably worse.
        For higher dimensions the problem becomes substantially harder and the differences between
        algorithms become less pronounced.
    }
    \label{fig:realgap}
\end{figure}

To better assess the quality of the solutions obtained by the different algorithms, we compute a \emph{real optimality gap} using the best lower bound obtained across all runs and configurations.
The resulting gaps are shown in Figure~\ref{fig:realgap} using a logarithmic color scale.

Interestingly, the real gaps are substantially smaller than the reported gaps. This indicates that the relatively large reported gaps in some runs are mainly due to weak lower bounds rather than poor feasible solutions.

In particular, the monolithic constraint programming approach often finds competitive solutions despite reporting weak bounds. Similarly, several incremental variants produce high-quality solutions even when the reported gap appears large.

The comparison between sampling strategies remains consistent with previous observations. The RAND strategy performs significantly worse than the metric-based alternatives, while ECC generally provides the most reliable performance.

For dimension $3$, the internal time limit strategy again appears beneficial. In contrast, for dimension $5$, the differences between algorithms become smaller and the problem becomes intrinsically harder. Even the best configurations produce real gaps of a few percent in this setting.

These results suggest that while the incremental algorithm effectively addresses the dependence on the number of points, the dimension of the space remains the main source of computational difficulty. At higher dimensions, the points become sparse and the distance between any two points becomes hard to differentiate, complicating the identification border points. This so-called \emph{curse of dimensionality} explains that all sampling metrics provide similar results and very high gaps.

\subsubsection*{Performance profiles}

\begin{figure}
    \centering
    \includegraphics[width=0.8\textwidth]{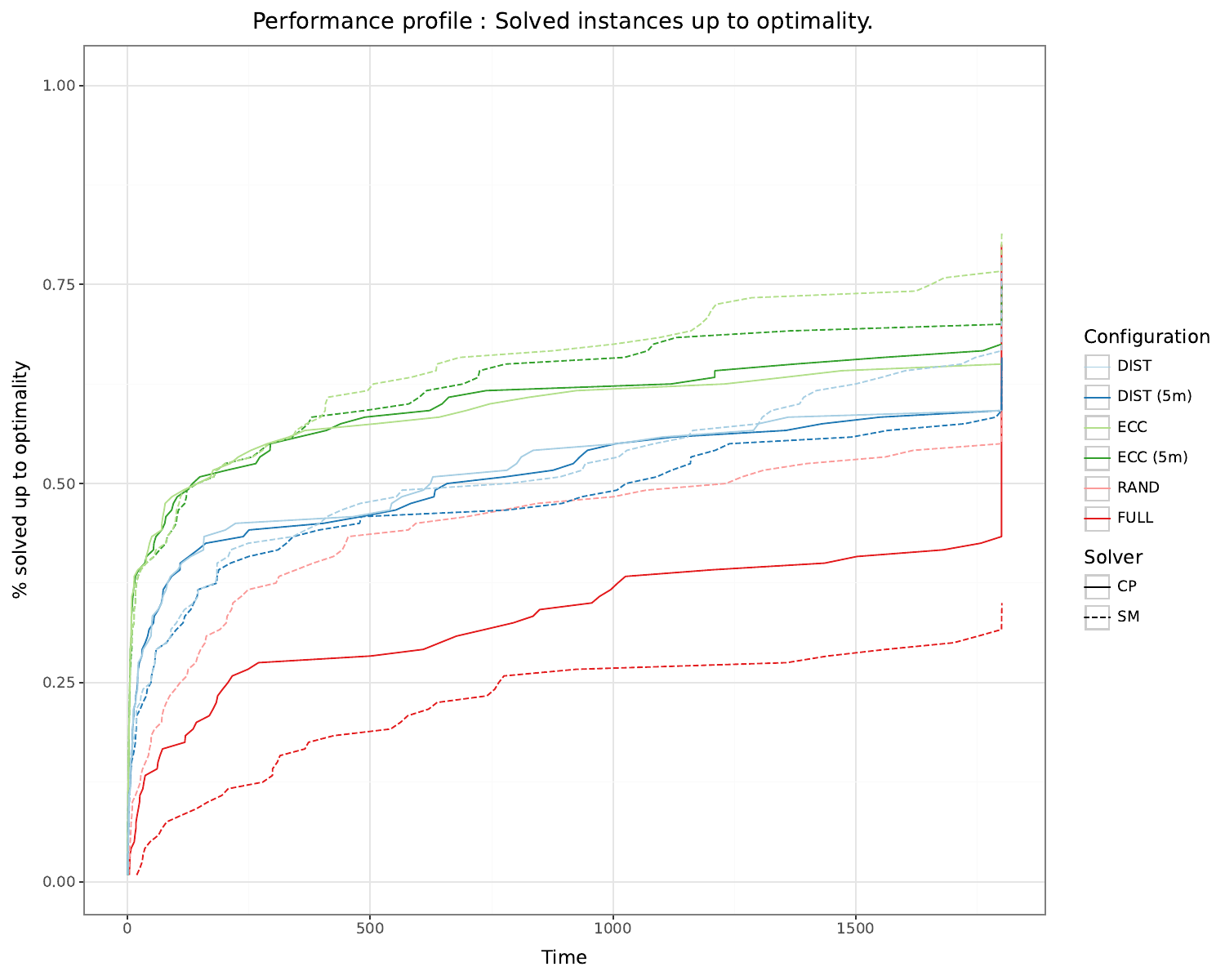}
    % \caption{Performance profiles measuring the percentage of instances solved to optimality as a function of time.}
    \caption{
        Performance profiles showing the fraction of instances solved to optimality within a given
        time limit.
        The curves represent the percentage of instances solved as a function of time for the
        different algorithmic configurations.
        Incremental algorithms clearly dominate the monolithic approach.
        Among the sampling strategies, ECC consistently solves the largest number of instances,
        followed by DIST, while the random sampling strategy (RAND) performs significantly worse.
    }
    \label{fig:perfprofile}
\end{figure}

Finally, Figure~\ref{fig:perfprofile} presents performance profiles measuring the percentage of instances solved to optimality as a function of time.
The results clearly show that incremental algorithms dominate the monolithic approach. Among the incremental sampling metric variants, the performance ordering is consistent across the experiments:

\[
	\text{ECC}  \succ \text{DIST}  \succ \text{RAND}.
\]
The ECC strategy consistently solves the largest number of instances within the time limit, confirming its robustness across different instance sizes and dimensions.

\section{Conclusions and further remarks}\label{sec:conclusions}

In this work we proposed an incremental exact approach for the hyper-rectangular
clustering problem with axis-parallel clusters (\prob).
Existing exact approaches for \prob{} typically rely on solving integer programming
formulations directly, which limits their applicability to relatively small
instances. The main idea of our approach is to exploit the fact that optimal
cluster boundaries are determined by a small subset of points.
By iteratively solving the problem on subsets of the data and enlarging these
subsets only when necessary, the algorithm is able to certify optimality for the
original instance while solving significantly smaller subproblems.

Our computational experiments show that this strategy substantially improves
scalability with respect to the number of points.
While the monolithic formulation quickly becomes intractable as the instance size
grows, the incremental algorithm remains able to solve instances with up to
10.000 points in many cases.
Moreover, even when optimality cannot be proven within the time limit, the
algorithm consistently produces high-quality solutions with very small optimality
gaps.

The experiments also highlight the importance of the sampling strategy used to
select candidate points.
Among the tested alternatives, the eccentricity-based metric ECC clearly
provides the most reliable performance, consistently outperforming both the
distance-based metric and random sampling.
These results suggest that identifying points that are likely to lie on cluster
boundaries is a key ingredient for the effectiveness of the incremental framework.

Finally, our experiments reveal an interesting structural property of the problem.
While the incremental approach effectively mitigates the dependence on the number
of points, the dimension of the space remains the main source of computational
difficulty.
Instances with higher dimensionality become significantly harder, even when the
number of points is moderate, indicating a strong curse-of-dimensionality effect
in \prob.

Several directions for future research remain open.
First, other sampling strategies could be investigated to further improve the
identification of boundary-defining points.
Second, alternative formulations or solution techniques could be explored for
the subproblems arising in the incremental framework.
Finally, it may be interesting to investigate variants of the algorithm in which
points can also be removed from the working subset, rather than only added,
potentially leading to even smaller subproblems and faster convergence.

% \printbibliography

\section*{Acknowledgements}

This work was partially funded by \emph{CY Generations}, a program supported by the French government
grant: \emph{``Investissements d\textquotesingle avenir'' \#France2030}.

\bibliographystyle{elsarticle-harv}
\bibliography{hrc}

@article{agarwal2005geometric,
  title={Geometric approximation via coresets},
  author={Agarwal, Pankaj K and Har-Peled, Sariel and Varadarajan, Kasturi R and others},
  journal={Combinatorial and computational geometry},
  volume={52},
  number={1},
  pages={1--30},
  year={2005}
}

@inproceedings{feldman2011unified,
  title={A unified framework for approximating and clustering data},
  author={Feldman, Dan and Langberg, Michael},
  booktitle={Proceedings of the forty-third annual ACM symposium on Theory of computing},
  pages={569--578},
  year={2011}, 
  doi={10.1145/1993636.1993712}
}

@inproceedings{badoiu2002approximate,
  title={Approximate clustering via core-sets},
  author={B{\=a}doiu, Mihai and Har-Peled, Sariel and Indyk, Piotr},
  booktitle={Proceedings of the thiry-fourth annual ACM symposium on Theory of computing},
  pages={250--257},
  year={2002}, 
  doi={10.1145/509907.509947}
}

@inproceedings{dbscan,
  title={A density-based algorithm for discovering clusters in large spatial databases with noise.},
  author={Ester, Martin and Kriegel, Hans-Peter and Sander, Jorg and Xu, Xiaowei and others},
  booktitle={Knowledge Discovery and Data Mining},
  volume={96},
  number={34},
  pages={226--231},
  year={1996}
}

@article{clustering_bap,
  title={A branch-and-price algorithm for the hyper-rectangular clustering problem with axis-parallel clusters and outliers},
  author={Delle Donne, Diego and Marenco, Javier},
  journal={Computational Optimization and Applications},
  volume={90},
  number={2},
  pages={583--605},
  year={2025},
  publisher={Springer},
  doi={10.1007/s10589-024-00637-w}
}

@article{cplex,
	title={User’s Manual for CPLEX},
	author={{Cplex, IBM ILOG}},
	journal={International Business Machines Corporation},
	year={2022}
}

@article{marenco23,
	title = {An integer programming approach for the hyper-rectangular clustering problem with axis-parallel clusters and outliers},
	journal = {Discrete Applied Mathematics},
	volume = {341},
	pages = {180-195},
	year = {2023},
	issn = {0166-218X},
	doi = {10.1016/j.dam.2023.08.004},
	author = {Javier Marenco}
}

@inproceedings{mago12,
  title={Classification with axis-aligned rectangular boundaries},
  author={Park, Sung Hee},
  booktitle={Cross-Disciplinary Applications of Artificial Intelligence and Pattern Recognition: Advancing Technologies},
  pages={355--366},
  year={2012},
  publisher={IGI Global Scientific Publishing},
  doi={10.4018/978-1-61350-429-1.ch019}
}

@techreport{park09,
	author={Sung Hee Park and Jae-Young Kim},
	title={Unsupervised clustering with axis-aligned rectangular regions},
	institution={Stanford University},
	year={2009},
	type={Technical Report}
}

@misc{cpsatlp,
title = {{CP-SAT} v9.12},
author = {Laurent Perron and Fréderic Didier},
month = {February},
year = 2025,
howpublished = {\url{https://developers.google.com/optimization/cp/cp_solver/}},
note = {Google}
}

@misc{ortools,
title = {{OR-Tools} v9.12},
author = {Laurent Perron and Vincent Furnon},
month = {February},
year = 2025,
howpublished = {\url{https://developers.google.com/optimization/}},
note = {Google}
}
\end{document}